\documentclass[%
 reprint,
 amsmath,amssymb,
 aps,
]{revtex4-1}

\usepackage{graphicx}
\usepackage{dcolumn}
\usepackage{bm}

\begin{document}

\preprint{APS/123-QED}

\title{Neutron EDM in Four Generation Standard Model}

\author{Junji Hisano$^{a,b}$, Wei-Shu Hou$^{c,d}$, Fanrong Xu$^{c}$}
 \affiliation{
$^{a}$Department of Physics, Nagoya University, Nagoya 464-8602, Japan\\
$^{b}$IPMU, TODIAS, University of Tokyo, Kashiwa, 277-8583, Japan\\
$^{c}$Department of Physics, National Taiwan University, Taipei, Taiwan 10617\\
$^{d}$National Center for Theoretical Sciences, National Taiwan University, Taipei, Taiwan 10617}


\begin{abstract}
A fourth generation of quarks, if it exists, may provide
sufficient $CP$ violation for the baryon asymmetry of the Universe.
We estimate the neutron electric dipole moment in the presence of a
fourth generation, and find it would be dominated by the strange
quark chromoelectric dipole moment, assuming it does not get wiped
out by a Peccei-Quinn symmetry. Both the three electroweak loop
and the two-loop electroweak/one-loop gluonic contributions are considered.
With $m_{b'}$, $m_{t'}$ at 500 GeV or so that can be covered at the LHC,
and with a Jarlskog CPV factor that is consistent with hints of New Physics
in $b\to s$ transitions, the neutron EDM is found around $10^{-31}\ e\,$cm,
still far below the $10^{-28}\ e\,$cm reach of the new experiments being
planned or under construction.
\begin{description}
\item[PACS numbers]
 13.40.Em 
 11.30.Er 
 14.65.Jk 
\end{description}
\end{abstract}

\pacs{Valid PACS appear here}
\maketitle


\section{\label{sec:Intro}INTRODUCTION and Motivation\protect\\}

The Kobayashi--Maskawa (KM) mechanism~\cite{KM} for $CP$ violation
(CPV) has been verified by the Belle and BaBar experiments~\cite{PDG}.
Constituting the flavor and CPV part of the Standard Model (SM),
it falls short of the observed Baryon Asymmetry of the Universe (BAU)
by many orders of magnitude. However, it was pointed out that,
by extending to four quark generations~\cite{4S4G}, SM4,
the KM picture may have enough~\cite{HouCJP} CPV phase for BAU.
The strength of phase transition, needed to satisfy the third Sakharov condition,
i.e. departure from equilibrium, remains an issue.
But interest has renewed~\cite{CDFCMS} in
the direct search of fourth generation quarks at hadron colliders,
where the LHC should finally be able to discover, or rule out
once and for all~\cite{AH06}, fourth generation quarks.

The long quest for neutron electric dipole moment (nEDM) has been
motivated by BAU, as the latter implies the existence of
new CPV sources beyond SM. Given the large jump in CPV,
it is of interest to ask what nEDM value one might expect for SM4.
The KM mechanism cleverly shields itself from nEDM.
At the one weak loop level, the CKM factors come always conjugate
to each other so the CPV phase cancels.
It was shown~\cite{Shabalin} by Shabalin, at two loop in SM,
that the sum over all diagrams for single quark
electric dipole moments (qEDM) vanish.
It was then shown that, bringing in a further gluon loop
(two-loop electroweak/one-loop gluonic) breaks the identical cancelation,
leading to $d_d \sim 10^{-34}\;e\,$cm~\cite{Khriplovich,CzKr97}.
Considerations of long distance (LD) effects allow a value of $d_n$
that is two orders of magnitude higher~\cite{KhZh82}.

The current limit for nEDM, $2.9 \times 10^{-26}\;e\,$cm
at 90\% C.L.~\cite{Baker06}, is from
the RAL-Sussex-ILL experiment which operated at Grenoble.
Compared with dropping an order of magnitude per decade~\cite{LG09}
since the 1950s, the pace has slowed.
The chief limiting factor is the number of neutrons.
There is, however, a renewed effort, by
the CryoEDM collaboration at Grenoble,
the nEDM collaboration at PSI, and
the nEDM collaboration at the Spallation Neutron Source
(SNS) at Oak Ridge National Lab,
to push~\cite{LG09, expEDM} first towards ${\cal O}(10^{-27})\;e\,$cm,
then eventually down to $10^{-28}\;e\,$cm.
There are efforts also at J-PARC~\cite{J-PARC}, and TRIUMF~\cite{Tri},
with similar aims.
All these efforts use ultra cold neutrons (UCN),
hence many more usable neutrons,
together with stronger electric fields, and better magnetic field control.
The SNS experiment is the most innovative,
injecting polarized $^3$He into liquid $^4$He.
The $^3$He serves as both~\cite{LG09} ``co-magnetometer"
and as detector for neutron capture,
while the liquid $^4$He serves as scintillator,
as well as a superthermal source for UCN,
and it is much more tolerant of a higher electric field.

These new ambitious experiments are motivated in part by the
cosmological limit of $10^{-28}\;e\,$cm, if supersymmetry (SUSY) is
relevant~\cite{CLPR-M} for BAU.
What is the situation for SM4, the Standard Model with four quark generations,
given that CPV is greatly enhanced? In this paper we address this issue.

The nEDM for SM4 was already considered in the 1990s~\cite{HP95, HP96, XGHe}.
However, they went largely unnoticed because
the fourth generation fell out of favor in a similar time frame,
by the twin issues of neutrino counting and
electroweak precision tests~\cite{PDG, 4S4G}.
But since the discovery of atmospheric neutrinos,
we know the neutrinos have mass, which calls for
New Physics beyond SM.
The fourth generation neutral lepton~\cite{4S4G},
which does not enter our discussion, if it exists, must be heavy.
More recently, it was pointed out~\cite{Kribs07} that
the electroweak $S$ and $T$ variables do allow a somewhat split
(less than $M_W$) but close to degenerate fourth generation, if
one allows the Higgs boson to be heavier at the same time.

In the following sections, we first collect some relevant
formulas, then proceed to make a numeric estimate.
We end with some discussions and a conclusion.

\section{\label{sec:Formulas}Some Relevant Formulas\protect\\}

Starting from the effective Lagrangian of all CPV operators
up to dimension 5, the neutron EDM was evaluated in the
QCD sum rule framework. In terms of quark
EDMs and chromo-EDMs (CEDM), one has~\cite{PR01}
\begin{eqnarray}
d_n = (0.4 &\pm& 0.2)
     \Biggl[ \chi m_{*}(4e_d - e_u)\left(\bar{\theta}
          - \frac{1}{2} m_0^2\frac{\tilde{d}_s}{m_s}\right) \nonumber\\
    & + &
          \frac{1}{2}\chi m_0^2 \left(\tilde{d}_d - \tilde{d}_u\right)
          \frac{4e_dm_d + e_um_u}{m_u + m_d}
         \nonumber\\
    & + & \frac{1}{8}\left(4\tilde{d}_d\alpha^+_d
                           - \tilde{d}_u\alpha_u^+\right)
         + (4d_d - d_u)\Biggr],
\end{eqnarray}
where $1/m_{*} = 1/m_u + 1/m_d + 1/m_s \cong 1/m_u + 1/m_d$,
$\bar{\theta}=\sum_q \theta_q+\theta_G$ is the combined quark and gluonic
$\theta$ term, $\alpha^\pm_q = e_q(2\kappa \pm \xi)$, and $\chi$, $m_0^2$,
$\kappa$ and $\xi$ are condensate susceptibilities.
The large factor of 3 uncertainty inherent in the overall $0.4 \pm 0.2$
coefficient reflects the large hadronic uncertainty,
as determined in the sum rule approach. Thus,
our estimates that follow are only aimed at the order of magnitude.

The interesting subtlety is that, when a Peccei-Quinn symmetry~\cite{PQ}
is invoked to remove the $\bar\theta$ term (setting it to zero), it
induces additional CPV terms~\cite{BiUr91} to the axion potential that
completely cancels the strange quark CEDM (sCEDM) contribution~\cite{PR00},
independent of details of the axion potential. While remarkable,
as we shall see, the sCEDM is of the greatest interest in SM4.
Furthermore, three decades of axion search has so far come to naught.
Given that there are models of spontaneous CPV, such as the
Nelson-Barr mechanism~\cite{NeBa84}, where $\theta = 0$ while
the KM phase (phases for SM4 case) is generated by spontaneous symmetry
breaking, we shall ignore the $\bar\theta$ term while keeping the
qCEDM terms.

We follow Ref.~\cite{PR01} and take the numerical values
$m_0^2 = 0.8\ \mathrm{GeV}^2$, $\chi = -5.7 \pm 0.6\ \mathrm{GeV}^{-2}$,
 $\xi=-0.74 \pm 0.2$, and $\kappa=-0.34 \pm 0.1$.
Numerically, we then have
\begin{eqnarray}
d_n &=& (0.4\pm0.2)\Bigl[
          1.9 \times 10^{-16}\, \bar{\theta}\;e\,\mathrm{cm} - 0.08e\,\tilde{d}_s
        \nonumber\\
    & & \ \ \ \ \ \ \ \ \ \ \ \
     +1.8 e\,\tilde{d}_d -1.4e\,\tilde{d}_u + (4d_d-d_u)\Bigr].
 \label{no-PQ}
\end{eqnarray}
The strange quark CEDM entered by $m_s$ being light enough,
such that it partakes in the $\bar qq$ condensation.

Analyzing the flavor structure of a typical three loop diagram shows
why the strange CEDM is highlighted, despite a smaller coefficient in
Eq.~(\ref{no-PQ}). A typical three loop diagram involves two nonoverlapping
$W$ boson loops, with one $Z$ or gluon loop. Following the quark line,
the $f$ quark (C)EDM has the following form~\cite{HP95, Posp94}
\begin{eqnarray}
& & i\,\sum_{j,k,l}\mathrm{Im}\,(V^*_{jf}V_{jk}V^*_{lk}V_{lf})\,fjk\,lf \nonumber\\
 =\ && \frac{i}{2} \sum_{j,k,l}\mathrm{Im}\,(V^*_{jf}V_{jk}V^*_{lk}V_{lf})\,f(jk\,l-lkj)f,
 \label{fjklf}
\end{eqnarray}
where $f$, $j$, $k$, $l$ stand for both flavor indices and the corresponding
Green function. The antisymmetry in Eq.~(\ref{fjklf}) is at the root of
Shabalin's argument.

Since we shall consider rather heavy $t'$ and $b'$
quarks in the loop, typical loop momenta would be at these large values.
Therefore, one can take $c = u \equiv \textsf{u}$, $d = s =b \equiv \textsf{d}$ as all effectively massless in loop propagators.
One then easily sees that the (C)EDM of the $u$ quark vanishes.
That is, performing the sum over $j$ and $l$ in
Eq.~(\ref{fjklf}) for $f = u$, using the unitarity relation
 $V^*_{ud}V_{kd} + V^*_{us}V_{ks} + V^*_{ub}V_{kb} = \delta_{uk} - V^*_{ub'}V_{kb'}$
 and the ``degeneracy" in mass for the $d$, $s$, $b$ propagators, one gets
\begin{eqnarray}
&i& \,\sum_{j,k,l}\mathrm{Im}(V^*_{uj}V_{kj}V^*_{kl}V_{ul})\,ujk\,lu \nonumber\\
 =\ & \frac{i}{2} & \sum_{k}\mathrm{Im}(V^*_{ub'}V_{kb'}V^*_{kb'}V_{ub'})
  \,u(\textsf{d}k\,b'-b'k\,\textsf{d})u = 0,
 \label{ujklu}
\end{eqnarray}
as phases pairwise cancel. Effectively there are only two generations in the loop.

The case for $f = d$, $s$ is therefore more interesting.
By similar token, one has
\begin{eqnarray}
& i &\,\sum_{j,k,l}\mathrm{Im}\,(V^*_{uj}V_{kj}V^*_{kl}V_{ul})\,fjk\,lf \nonumber\\
 =\ &{i}& \, \mathrm{Im}(V^*_{tf}V_{tb}V^*_{t'b}V_{t'f})
  \,f \bigl[ t\,(\textsf{d}-b')\,t' - t'(\textsf{d}-b')\,t \nonumber\\
 && \ \ \ \ \ \ \ \ \ \ \ \ \ \ \ \ \ \ \ \ \ \ \
  +\ t'(\textsf{d}-b')\,\textsf{u} - \textsf{u}(\textsf{d}-b')\,t' \nonumber\\
 && \ \ \ \ \ \ \ \ \ \ \ \ \ \ \ \ \ \ \ \ \ \ \
  +\ \textsf{u}(\textsf{d}-b')\,t - t\,(\textsf{d}-b')\,\textsf{u}\bigr]f.
 \label{ds-fjklf}
\end{eqnarray}
In Eq.~(\ref{ds-fjklf}), we have spelled out the sum over $k$, after utilizing
CKM unitarity as before. The sum of Green function factors contain the
``degeneracy" of $c = u \equiv \textsf{u}$ and $d = s = b \equiv \textsf{d}$ in the loop.
But one should treat the CKM coefficient with care, where use has been made of
the rather good approximate relation~\cite{HP95}
\begin{equation}
\mathrm{Im}(V^*_{tf}V_{tb'}V^*_{t'b'}V_{t'f})
 \cong - \mathrm{Im}(V^*_{tf}V_{tb}V^*_{t'b}V_{t'f}) \equiv \mathcal{J}_f,
\end{equation}
which is a consequence of the smallness of $V_{ub}$ (assuming that other CKM
elements that enter are not much smaller). As noted in Ref.~\cite{AH09},
for $f = s$, this relation effectively means that the
CKM ``triangle" (degenerated from a quadrangle)
governing $b'\to s$ transitions have essentially the
same area as the $b\to s$ triangle.
We will turn to numerical analysis in the next section,
but we can already see that the CKM factor for $f = s$ is
much larger than for $f = d$, which is the reason why
we highlight the strange quark CEDM.

The $s$ quark CEDM arising from the two-$W$ loop plus one gluon loop diagram
was estimated~\cite{HP95} using the external field method, to double log accuracy and
in the large $N_C$ limit, with the result of
\begin{equation}
\tilde{d}^{(g)}_s = -\mathcal{J}_s \,m_s\frac{G_F}{\sqrt{2}}
\frac{\alpha_s\alpha_W}{(4\pi)^4}\frac{5N_c}{6}\frac{m_t^2}{M_W^2}\frac{1}{2!}
\log^2\left(\frac{m^2_{t'}}{m_t^2}\right).
 \label{WWg}
\end{equation}

Replacing the gluon by a $Z$ boson loop, one might think it should be
subdominant. But since there are rather large quark masses in the loop,
it implies that rather large loop momenta might be relevant.
The derivative coupling nature of the longitudinal $Z$ boson,
or equivalently, that the Goldstone boson couples to
the heavy quark masses hence is nondecoupled, effectively
voids the above intuition.
By an ingenious argument of limiting to large loop momenta
and involving longitudinal vector bosons,
the authors of Ref.~\cite{HP96} were able to reduce the three-loop calculation
effectively to calculating three one-loop integrals, and the core of it is
an effective $i\to fZ$ transition involving the heavy fourth generation quark
in the loop. This is the familiar $Z$ penguin~\cite{IL80,HWS86},
and indeed it has been found~\cite{AH06} that
$b'\to bZ$ and $b'\to bg$ transitions are not too different in strength.
The upshot of the estimate (with the brutality of setting all logarithms
to order 1) of Ref.~\cite{HP96}, done in 1996, is
\begin{equation}
\tilde{d}^{(Z)}_s = -\mathcal{J}_s\,m_s\frac{G_F}{\sqrt{2}}\frac{\alpha^2_W}{(4\pi)^4}
\frac{m^2_t m^2_{t'}}{4M^4_{W}}\log\left(\frac{m^2_{t'}}{m^2_t}\right).
 \label{WWZ}
\end{equation}

Comparing Eqs.~(\ref{WWg}) and (\ref{WWZ}), aside from the double versus
single logarithm, one can see from
$\alpha_W/M^2_W = \sqrt{2}G_F/\pi = 1/\pi v^2$
that one is comparing $5N_c\alpha_s/6$ with $\lambda_{t'}^2/4\pi$.
The gluonic effect is enhanced by the color factor, but the Yukawa coupling
grows with $m_{t'}^2$. Compared literally, they are actually comparable.
On the other hand,
in arriving at Eq.~(\ref{WWZ}), one has set all logarithms to 1. In this spirit,
both the double log (including the $1/2!$) in Eq.~(\ref{WWg}) and the single log
in Eq.~(\ref{WWZ}) should be treated as order one. Then, the gluonic effect
would be subdominant to the $Z$ effect, for $t'$ and $b'$ masses of order
500 GeV (or higher), a nominal value used by Ref.~\cite{HP96}, and which
we shall use in the next section.

Given the roughness of these calculations, and the great difficulty
in calculating genuine three electroweak loop diagrams, we shall take the
estimate of Eq.~(\ref{WWZ}) for our subsequent numerics.

\section{\label{sec:Numeric}Numerical Estimate\protect\\}

We shall use $m_{t'} \simeq m_{b'} \simeq 500$ GeV as our nominal
fourth generation quark mass. Above this value, one would pass
through the unitarity bound~\cite{UB78}, and the Yukawa coupling starts to
turn nonperturbative. For the other quark masses, we take
$m_u = 2.5$ MeV, $m_d = 5$ MeV, $m_s = 100$ MeV, $m_c = 1.3$ GeV,
$m_b = 4.2$ GeV, and $m_t = 165.5$ GeV,
where heavy quark masses are in $\overline{\rm MS}$ scheme.
The light quark mass values were in fact implicit in Eq.~(\ref{no-PQ}).
For the CKM products ${\cal J}_s$ and ${\cal J}_d$, we take the
nominal fit~\cite{HM10} to flavor data performed for $m_{t'} \simeq 500$ GeV,
where $V_{t'b} \simeq -0.1$,  $V_{t's} \simeq -0.06\, e^{-i75^\circ}$,
and  $V_{t'd} \simeq -0.003\, e^{-i18^\circ}$,
we get
\begin{eqnarray}
{\cal J}_s &=& \mathrm{Im}(V^*_{ts}V_{tb}V^*_{t'b}V_{t's})
 \simeq 2.4\times 10^{-4},
  \label{Js} \\
{\cal J}_d &=& \mathrm{Im}(V^*_{td}V_{tb}V^*_{t'b}V_{t'd})
 \simeq 1.7\times 10^{-7}.
  \label{Jd}
\end{eqnarray}
Note that ${\cal J}_s$ could be measured~\cite{HKX11}
in the next two years at the LHC,
but ${\cal J}_d$ would be harder to disentangle.

Although the range of uncertainty is large, it is clear that
$|{\cal J}_s| \gg |{\cal J}_d|$, which correlates with the fact
that $b\to d$ transitions, including in $B_d$ mixing, show little
sign of deviation from SM expectations, while for $b\to s$ transitions,
and especially in $B_s$ mixing, we have several indications for
sizable deviations from three generation SM. Note that the study of
Ref.~\cite{HM10} predated the summer Tevatron update on $\sin2\Phi_{B_s}$,
the CPV phase of the $\bar B_s$ to $B_s$ mixing amplitude, and
predicted a value lower than the $-0.5$ to $-0.7$ value given~\cite{HNS07}
in 2007 for $m_{t'} \simeq 300$ GeV.
In any case, for our purpose, we see from Eqs.~(\ref{Js}) and (\ref{Jd})
that, despite the smaller coefficient in Eq.~(\ref{no-PQ}), the
sCEDM is the most important quark contribution to $d_n$, because of
the CKM factor.

Putting in numbers, we find from Eq.~(\ref{WWZ}) that
\begin{equation}
\tilde d_s^{(4)} \simeq -4 \times 10^{-16}\ {\rm GeV}^{-1}
 \simeq -0.8 \times 10^{-29}\ {\rm cm},
  \label{sCEDM4}
\end{equation}
where the $W$-$W$-$g$ 3-loop effect of Eq.~(\ref{WWg}) is
treated as subdominant.
Treating the sCEDM as the leading effect in Eq.~(\ref{no-PQ}), then
\begin{equation}
 d_n^{(4)} =  (2.2 \pm 1.1) \times 10^{-31}\ e\, {\rm cm},
  \label{dn4}
\end{equation}
where the superscript indicates that this is the
estimated effect from the fourth generation.

\section{\label{sec:Discussion}Discussion and Conclusion\protect\\}

Though the estimate of Eq.~(\ref{dn4}) is considerably
larger than the SM result with three quark generations,
be it the quark level~\cite{CzKr97},
or hadronically enhanced~\cite{KhZh82},
it is very far from the sensitivities of next generation
experiments~\cite{LG09, expEDM, J-PARC, Tri},
nominally at $10^{-28}\ e\,{\rm cm}$.
What is worse, or perhaps intriguing, is that the
sCEDM effect of Eq.~(\ref{sCEDM4}) is at the mercy
of the Peccei-Quinn symmetry.
If a PQ symmetry is operative in Nature, then
the sCEDM effect is precisely canceled~\cite{BiUr91,PR00}.
In this case, one has a reduced formula. Rather than
Eq.~(\ref{no-PQ}), one gets~\cite{PR01}
\begin{eqnarray}
d_n^{\rm PQ} = (0.4 &\pm& 0.2)\Bigl[
        1.6e(2\tilde{d}_d + \tilde{d}_u) 
     + \; (4d_d - d_u)\Bigr],
 \label{PQ}
\end{eqnarray}
i.e. only dependent on the naive constituents of the neutron.
Note that the axion potential has also modified the $d$ and $u$
quark CEDMs, as it brings in analogous terms to the one
that canceled away the sCEDM effect.

From Eq.~(\ref{ujklu}), $d_u$ and $\tilde d_u$ vanish.
Thus, we need only consider $d_d$ and $\tilde d_d$ for SM4.
Assuming again that the analogue of Eq.~(\ref{WWZ})
dominates over the gluonic counterpart,
we use the formulas of Ref.~\cite{PR01}
for $d_s$ and Eq.~(\ref{WWZ}) to obtain $d_d$ and $\tilde d_d$
by simply shifting the CKM index, i.e. shifting from ${\cal J}_s$
of Eq.~(\ref{Js}) to ${\cal J}_d$ of Eq.~(\ref{Jd}),
and replacing $m_s$ by $m_d$.
We find
\begin{equation}
\tilde d_d^{(4)} \simeq -3\times 10^{-34}\ {\rm cm},\ \
d_d^{(4)} \simeq -4\times 10^{-34}\ e\,{\rm cm},
\end{equation}
and, from Eq.~(\ref{PQ}),
\begin{eqnarray}
d_n^{(4){\rm PQ}} = -(1 &\pm& 0.5) \times 10^{-33}\ e\,{\rm cm},
 \label{dn4PQ}
\end{eqnarray}
where $d_d$ contributes roughly twice as $\tilde d_d$.
These should be taken as very rough estimates.
Note that, unlike Eq.~(\ref{Js}), it
would take some time to refine Eq.~(\ref{Jd}).

We see that $d_d^{(4)}$ is stronger than in SM~\cite{CzKr97},
while $d_n^{(4){\rm PQ}}$, as estimated through sum rules,
is at $10^{-33}\ e\,{\rm cm}$ level.
Other LD effects might bring about another~\cite{KhZh82}
order (or maybe two) of magnitude enhancement.
But it should be clear that, with PQ symmetry operative,
even with four quark generations, nEDM is way below the
sensitivity of the next generation of experiments.

But what about LD effects to the sCEDM-driven result of Eq.~(\ref{dn4}),
without PQ symmetry?
Operators beyond dimension 5, such as dimension 6 operators involving
two quarks in the neutron, are beyond the scope of our investigation.
Even with $WWg$ loops, Eq.~(\ref{WWg}), of similar order of magnitude
and constructive with the $WWZ$ loop result of Eq.~(\ref{WWZ}) (hence
Eq.~(\ref{sCEDM4})), $d_n^{(4)}$ without PQ symmetry
still seems a couple of orders below $10^{-28}\ e\,{\rm cm}$.

One could also estimate from $s$ to $d$ transition elements induced
by gluon or $Z$, enhanced by pion-nucleon coupling at long distance.
It is possible~\cite{HNS05} that $K_L \to \pi^0\bar\nu\nu$
gets enhanced by a factor of 100 in SM4,
which means a factor of 10 at amplitude level. Given the two orders
of magnitude LD enhancement in SM~\cite{KhZh82},
it could be brought up by another order of magnitude in SM4,
hence to $10^{-31}\ e\,{\rm cm}$.
This is rather similar to the result in Eq.~(\ref{dn4}).
We remark that, one way to probe the sEDM and sCEDM effects,
even if PQ symmetry is operative, would be to measure
hyperon EDM, which is very interesting in its own right.
A rough estimate, by extrapolating from Eq.~(\ref{PQ}),
gives the order at $10^{-29}$--$10^{-28}\ e\,{\rm cm}$.
But if $10^{-28}$--$10^{-27}\ e\,{\rm cm}$ is still
the challenge of the next decade for neutron EDM
experiments, the measurement of the EDM of the
less abundant, shorter-lived hyperons would be
still farther away.

The CMS experiment has recently stated that
``data exclude SM4 Higgs boson with mass between
120 and 600 GeV at 95\% C.L.", while the $t'$ quark
is not seen below 450 GeV~\cite{Tonelli}. One may be entering
the strong coupling, beyond the unitarity bound regime
of $m_{t'} > 500$ GeV, with an associated composite,
massive (hence broad) ``Higgs" boson. In this regime,
the short distance results of Eqs.~(\ref{WWg}) and
(\ref{WWZ}) start to fail even as rough estimates,
as the loop functions themselves turn nonperturbative.

In conclusion, with four quark generations and with
Peccei-Quinn symmetry operative, the neutron EDM is
slightly enhanced above the SM value, but no more than
an order of magnitude, hence much below the sensitivities
of the next generation of experiments.
If PQ symmetry is absent, then a large enhancement is
possible through the $s$ quark CEDM,
which is correlated with possible effects in $b\to s$ transitions
that are of current interest.
However, it is still unlikely that the neutron EDM could reach
the $10^{-28}\ e\,{\rm cm}$ level sensitivity that may be
probed during the next decade.

\vskip0.3cm
\noindent{\bf Acknowledgement}.  The work of JH is supported
by Grant-in-Aid for Scientific research from the Ministry of
Education, Science, Sports, and Culture (MEXT), Japan, No. 20244037,
No. 20540252, and No. 22244021 (J.H.), and also by World Premier
International Research Center Initiative 14 (WPI Initiative), MEXT,
Japan.  WSH thanks the National Science
Council for an Academic Summit grant, NSC 99-2745-M-002-002-ASP, and
FX is supported under NTU grant 10R40044.

\end{document}